# FPGA Implementation of Short Critical Path CORDIC-Based Approximation of the Eight-Point DCT

Maxim Vashkevich [1)], Marek Parfieniuk [2)], Alexander Petrovsky [1)]

1) Belarusian State University of Informatics and Radioelectronics, 6 P. Brovky St., 220013, Minsk, Belarus, e-mail: vashkevich.m@gmail.com

2) Bialystok Technical University, ul. Wiejska 45A, 15-351 Bialystok, Poland, e-mail: marek@wi.pb.edu.pl

***Abstract:*** *This paper presents an efficient approach for multiplierless implementation for eight-point DCT approximation, which based on coordinate rotation digital computer (CORDIC) algorithm. The main design objective is to make critical path of corresponding circuits shorter and reduce the combinational delay of proposed scheme.*

## 1. INTRODUCTION

It is well know that the discrete cosine transform (DCT) has been widely used in many areas such as speech and image coding. In particular, the two-dimensional (2-D) DCT has been adopted in some international standards such as MPEG, JPEG and CCITT [1]. A 2-D DCT can be obtained by applying 1-D DCT over the rows followed by a 1-D DCT over the columns of the 8x8 data block [2]. Therefore the efficient implementation of DCT has become the most important issue in developing real-time embedded system. In mobile multimedia devices such as digital cameras, cell phone or pocket PCs hardware complexity as well as power consumption has to be minimized. To do this the great number of fast DCT algorithm were proposed [3], among which Loeffler algorithm [4] gained the lower bound of multiplicative complexity for 8-point DCT. It's required only 11 multiplication and 29 addition. But the common disadvantage of all fast DCT algorithm is that they still need floating point multiplication. These operations are very slow in software implementation and require large area and power in hardware. And therefore can not be used in mobile multimedia devises. So there is still the need to look for new design of DCT algorithm compromises better suited to particular application.

Mathematically, fast DCT is composed of additions and multiplications by constants. When implemented in hardware, the multiplication by constants are often implemented by a sequence of additions and shifts which is less expensive in terms of chip area and power consumption [5]. These implementations of transforms are referred to as *multiplierless*. The binDCT seems to be the most notable result in this field [6]. This transform is based on VLSI-friendly lattice structure and derived from DCT matrix factorization by replacing plane rotations with lifting schemes. Another popular way of multiplierless implementation of DCT is to use the coordinate rotation digital computer (CORDIC) algorithm [7]-[9]. Since the CORDIC algorithm leads to a very regular structure suitable for VLSI implementation.

In [10], it has been concluded that the length of the critical path, i.e. the maximum number of adders operating in cascade, strongly affects the performance of a hardware implementation of DCT. It has been shown that 30-40% decreases in delay and power consumption were obtained after shortening the critical path from 10 to 7, even through at the cost of increasing the total number of adders. In [7] were found that it is possible to optimize CORDIC-based structures to shorten the critical path to 5-6 additions still having good coding performance.

In this paper we discuss the FPGA implementation of CORDIC-based approximation of the eight-point DCT proposed in [7]. The paper begins with a review of fast Loeffler's algorithms and their multiplierless variant. Then the details of the proposed FPGA implementation scheme are given.

## 2. BASE STRUCTURE

In [7] starting point for derivation of short critical path approximation of 8-point DCT is signal flow graph of one of several possible Loeffler's algorithm (Fig. 1).

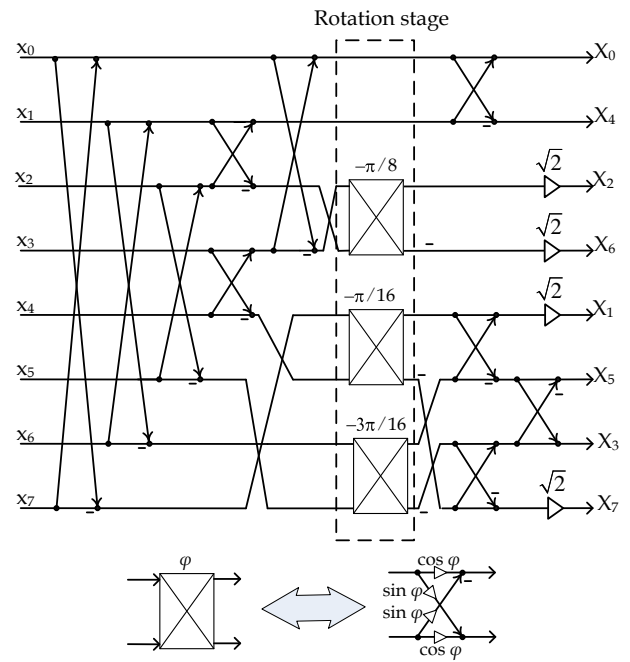

**Fig 1 – Signal flow graph of a Loeffler's algorithm.**

The only nontrivial operations are the three plane rotation by the angel: $\alpha = -\frac{1}{8}\pi$, $\beta = -\frac{1}{16}\pi$, $\gamma = -\frac{3}{16}\pi$ of the general form

$$R(\varphi) = \begin{bmatrix} \cos\varphi & -\sin\varphi \\ \sin\varphi & \cos\varphi \end{bmatrix}. \qquad (1)$$

In [6] using the lifting-based approach were showed that fast and accurate approximations of DCT can be obtained without using any multiplication. The obtained family of such transforms differing in accuracy and efficiency has been called the binDCT.

Another way of multiplierless implementation of plane rotation is CORDIC algorithm were also considered. In [11] special attention is given to constructing transform approximation maintaining orthogonality regardless of their coefficient quantization. On, contrary, in [10], performance maximization was of interest, especially from the hardware implementation point of view. In [7] were presented a novel family of CORDIC-based algorithms with short critical paths. Here we give the detail of FPGA implementation of DCT approximation algorithm proposed in [7].

## 3. CORDIC ALGORITHM

The CORDIC algorithms are an efficient method of producing a variety of trigonometric, hyperbolic and linear function [12].

In order to realized a plane rotation in CORDIC algorithm the rotation angle $\varphi$. is decomposed by the angle set called the CORDIC arc tangent radix (ATR) as follows [9]

$$\varphi = \sum_{i=0}^{M-1} \sigma_i \tan^{-1}(2^{-i}). \quad (2)$$

where $\sigma_i \in \{1,0,-1\}$. Then, the plane rotation is performed by the iterative equations given by

$$\begin{cases} x_{i+1} = x_i - \sigma_i 2^{-i} y_i \\ y_{i+1} = y_i + \sigma_i 2^{-i} x_i \end{cases} \quad 0 \le i \le M-1 \quad (3)$$

Note that each iteration assume the scaling of the vector, where the scale factor for the $i$th iteration is given by

$$k_i = \sqrt{1 + \sigma_i^2 2^{-2i}} \quad (4)$$

and the total scale factor $K$ is given by

$$K = \prod_{i=0}^{M-1} k_i. \quad (5)$$

Using (2)-(5) we can rewrite (1) in the following form

$$R(\varphi) \approx \frac{1}{K} \prod_{k=0}^{M-1} \begin{bmatrix} 1 & -\sigma_k 2^{-k} \\ \sigma_k 2^{-k} & 1 \end{bmatrix}. \quad (6)$$

Originally CORDIC algorithm allows $\sigma_i$ possessing the value $\pm 1$. However, recent papers show that computational savings can be achieved by allowing omitting and repeating some iteration. The elementary rotations also called microrotations.

The common approach of utilization CORDIC algorithm adjusted to DCT is to choose set of microrotations as close to required rotation angle as it possible. In contrast to this approach in [7] were proposed another method. The main difficulty that arises when we use CORDIC algorithm is necessity of implementation scaling. In order to extract scaling factors (for rotation by $\beta$ and $\gamma$) outside the transform core in [7] were decided to approximate angles $\beta$ and $\gamma$ with the same set of the absolute values of microrotation, thus scale factor for rotation by $\beta$ and $\gamma$ became equal and could be extracted outside the transform. There is no problem with scaling extraction for the rotation by $\alpha$.

It could also be noted that the scaling that require division by irrational numbers cannot be performed exactly using fixed-point arithmetic. However, in the most popular international standard such as MPEG and JPEG the DCT unit is followed by a quantizer, where DCT outputs are scaled by the pointwise division by the corresponding scaling constants that are stored in the quantization table [1]. Thus, each scaling factor of the DCT outputs can be incorporated into the corresponding scaling constant without requiring any additional hardware.

## 4. FPGA IMPLEMENTATION OF CORDIC-BASED DCT

Consider a variant C of approximation DCT given in [7]. In this case angles $\beta$ and $\gamma$ approximated with microrotation $i=\{1,2,4\}$, $\sigma^\beta_i = \{-1,1,1\}$, $\sigma^\gamma_i = \{-1,-1,1\}$ and for angle $\alpha$ $i=\{1,4\}$ and $\sigma^\alpha_i = \{-1,1\}$. The scheme obtained this way shown in Fig. 2.

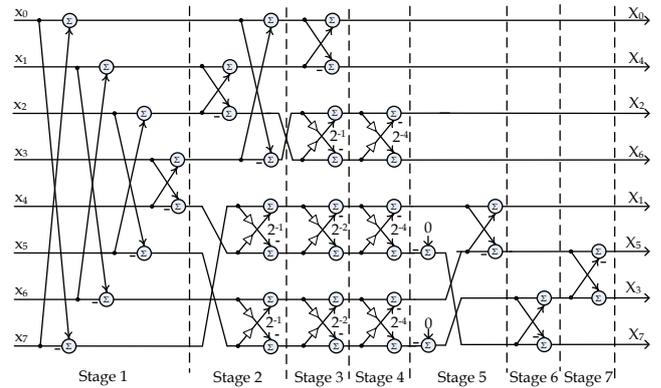

**Fig 2 – Signal flow graph of variant C of 8-point CORDIC-based DCT approximation (scaling factor omitted).**

It should be noted that practically critical path of the suggested scheme contain 7 adders. It is different from the result mentioned in [7], where critical path seems consist of 6 additions. Optional two adders appears in stage 5, where negation of two intermediate data sample presented in 2's complement code need to be implemented.

Examine stage 5 and stage 6 more detail. The first simplification that can be made is merging lower adder in stage 5 with adders in stage 6 (Fig.3)

It is known that inversion in 2's complement code performs as

$$-A = \bar{A} + 1 \quad (7)$$

where $\bar{A}$ it is simple bitwise inversion. Using (7) the second simplification we made by replacement of upper

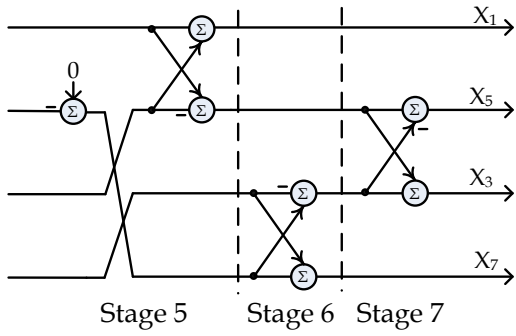

**Fig.3 – Merging of the adder from stage 5 to stage 6**

adder that perform negation (stage 5) with simple NOT gates (Fig. 4).

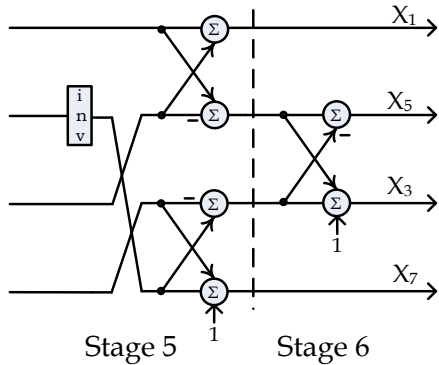

**Fig. 4 – Merging of stage 5 and 6.**

Addition of least significant bit (LSB) according to (7) can be made in the next stage (addition of LSB implements without extra hardware resources because conventional adder always have carry input port). However, for output $X_5$ LSB bit remain uncompensated.

Thus we lost one LSB of accuracy for one DCT output, but improving scheme performance. Flow graph obtained after proposed simplification depicted in Fig 5.

This scheme can be regarded as combinational circuit with total delay

$$T_{DCT} = 6T_{ADD} + T_{NOT} \qquad (8)$$

where $T_{ADD}$ - adder delay, and $T_{NOT}$ - delay of gate NOT

## 5. EXPERIMENTAL RESULT

The proposed method of approximation DCT realized with the FPGA place and route (PAR) process to determine the exact hardware cost. We use Xilinx Virtex series of FPGA for our experimentation. The hardware cost is measured as the total number of slices required to implement the design. A Xilinx Virtex slice contains two D-type flip-flops and two four-input lookup tables (LUT). As far as we implement a combinational circuit there is no flip-flops is needed. For more accurate measure we provide information about occupied slices and LUTs. Proposed solution implemented on FPGA XC4VLX25. Input data is 8-bit width, output data – 12 bit. Variant I of tested solution used only simplification pictured in Fig.3 , variant II is the scheme in Fig.5. In out experiments we have compared our solution with another low complexity DCT approximation algorithm binDCT of type C [6].

Table 1 shows hardware cost for DCT approximations. Table II compares the complexity and critical path of CORDIC-based approximation of DCT and binDCT.

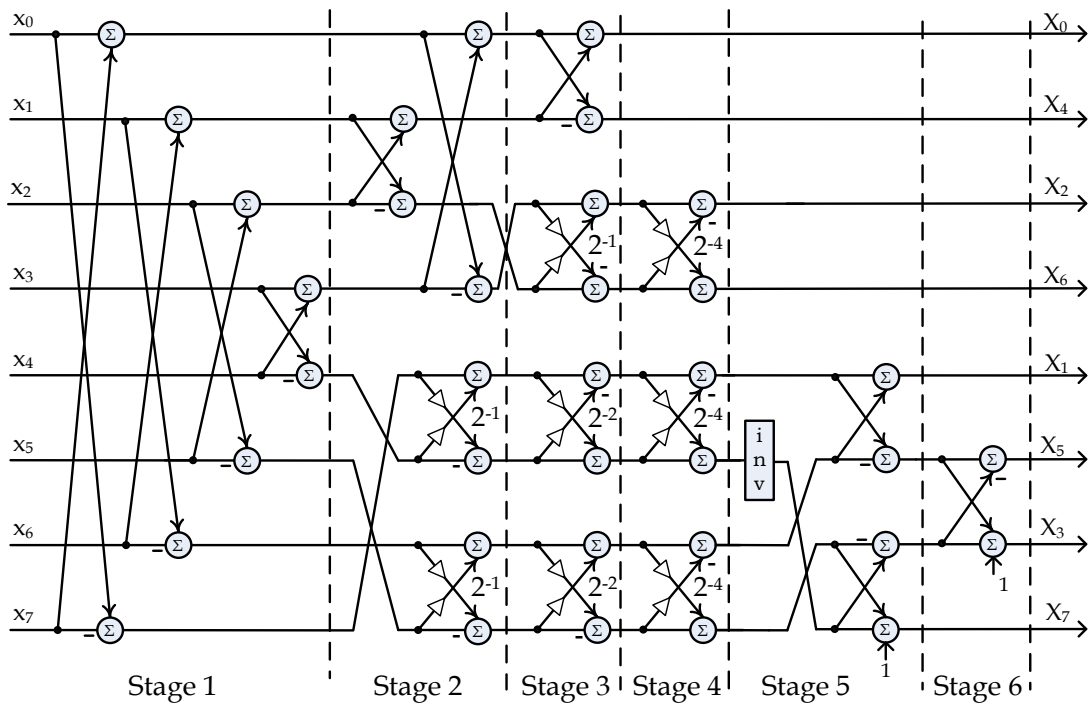

**Fig. 5 – Proposed signal flow graph for FPGA implementation.**

**Table 1. Hardware cost for DCT approximation**

| Parameters | Variant I | Variant II | binDCT-C |
|---|---|---|---|
| LUTs | 396 | 395 | 317 |
| SLICEs | 206 | 211 | 165 |

**Table 2. Arithmetical complexity and critical path of DCT approximation**

| Parameters | Variant I | Variant II | binDCT-C |
|---|---|---|---|
| No. of additions | 37 | 36 | 30 |
| No. of shifts | 16 | 16 | 12 |
| No. bit inversion | 0 | 1 | 0 |
| Critical path | $7T_{ADD}$ | $6T_{ADD} + T_{NOT}$ | $9T_{ADD}$ |

It can be noted that at expense of nearly 20 % in number of addition we can make critical path shorter in about 35%.

## 6. CONCLUSION

This paper has presented an efficient approach for 8-point CORDIC-based DCT approximation suitable for FPGA implementation. The proposed architecture requires 36 add, 16 shift and 1 bit inverse operations to carry out the DCT. Also critical path of given solution contain only 6 adder.

## 7. REFERENCES

[1] R.C. Gonzalez, R.E. Woods, *Digital Image Processing*, Prentice-Hall, Inc., New Jersey, 2001.

[2] S.Wolter, D. Birreck, R. Laur Classification for 2D-DCTs and a new architecture with distributed arithmetic, *Proceedings IEEE International Symposium on Circuits and Systems(ISCAS-1991)*, San Diego, USA, 11-14 Jun 1991, pp. 2204-2207.

[3] K. R. Rao, P. Yip, *Discrete Cosine Transform – Algorithms, Advantages, Applications*, Academic, New York, 1990.

[4] C. Loeffler, A. Ligtenberg, G.S. Moschyz Practical fast 1-D DCT algorithms with 11 multiplications, *Proceedings IEEE International Conference on Acoustics, Speech and Signal Processing (ICASSP-1989)*, Glasgow, Scotland, 23-26 May 1989, pp. 998-991.

[5] A. C. Zelinski, M. Puschel, S. Misra, J.C. Hoe Automatic cost minimization for multiplierless implementations of discrete signal transforms *Proceedings IEEE International Conference on Acoustics, Speech and Signal Processing (ICASSP-04)*, Pittsburgh, USA, 17-21 May 2004, pp. 221-224.

[6] T.D. Tran, The binDCT: fast multiplierless approximation of the DCT, *IEEE Processing Lett.*, vol. 7 no. 6, pp. 141-144, 2000.

[7] M. Parfieniuk, Shortening the critical path in CORDIC-based approximation of the eight-point DCT, *International Conference on Signal and Electronic Systems (ICSEC-2008)*, Krakow, Poland, September 14-17, pp. 405-408.

[8] B. Heyne, C.C. Sun, J. Goetze, S.J. Ruan, A computationally efficient high-quality CORDIC based DCT, *14th European Signal Processing Conference (Eusipco2006)*, Florence, Italy, 4-8 September 2006, pp. 11-15.

[9] S. Yu, E.E. Swartzlander Jr., A scaled DCT architecture with the CORDIC algorithm, *IEEE Transactions on Signal Processing*, vol.50, no. 1, pp. 160-167.

[10] C.-C. Sun, S.-J. Ruan, B. Heyne, and J. Goetze, Low-power and high-quality CORDIC-based Loeffler DCT for signal processing, *IET Circuits Devices Syst.*, vol. 1, no. 6, pp.623-655, 2007.

[11] M. Parfieniuk and A. Petrovsky, Structurally orthogonal finite precision implementation of the eight point DCT, *Proceedings IEEE International Conference on Acoustics, Speech and Signal Processing (ICASSP-06)*, Toulouse, France, 14-19 May 2006, pp. 936-939.

[12] J.E. Volder, The CORDIC trigonometric computing technique, *IRE Trans. Electron. Comput.*, vol. EC-8, pp. 330-334, 1959.